\documentstyle[preprint,eqsecnum,aps]{revtex}
\begin{document}
\draft
\preprint{\bf TSL/ISV-97-0173}
\title{Cross section and analyzing power of
\mbox{\boldmath $\vec{p}p\to pn\pi^+$}  near threshold}
\author{G\"oran F\"aldt\cite{gf}}
\address{Division of Nuclear Physics, Uppsala University, Box 535,
S-751 21 Uppsala, Sweden}
\author{Colin Wilkin\cite{cw}}
\address{University College London, London, WC1E 6BT, United Kingdom}
\date{\today}
\maketitle
\begin{abstract}
The cross section and analyzing power of the $\vec{p}p\to pn\pi^+$ reaction
near threshold are estimated in terms of data obtained from the $\vec{p}p\to
d\pi^+$ and $pp\to pp\pi^0$ reactions. A simple final state interaction
theory is developed which depends weakly upon the form of the pion-production
operator and includes some Coulomb  corrections. Within the uncertainties of
the model and the input data, the approach reproduces well the measured energy
dependence of the total cross section and the proton analyzing power at a 
fixed pion c.m. angle of $90^{\circ}$, from threshold to
 $T_p = 330$~MeV. The variation of the
differential cross section with pion angle is  also very encouraging.
\end{abstract}

\narrowtext

\protect\section{Introduction}

The $pp\to d\pi^+$ reaction has been measured for many years \cite{Arndt},
either for its own intrinsic physics interest or as a calibration for other
reactions. However, because of the importance of establishing the precise beam
energy, accurate experiments near threshold could be carried out much easier
with the advent of cooled proton beams \cite{pppid1,pppid2}. The region within
a few MeV of threshold is very interesting because it shows the interplay of 
the dominant $p$-wave production, induced by the $\Delta$ resonance, and 
the small $s$-wave rescattering terms. 

Since the phase-space for producing a three-body final state varies like $Q^2$,
where $Q$ is the c.m.\ kinetic energy in the exit channel, the determination of
the beam energy is even more crucial in such experiments near threshold. The
accurate measurements of the $pp\to pp\pi^0$ total cross section at the Indiana
\cite{Meyer} and CELSIUS storage rings \cite{Bondar} down to $Q\approx 1$~MeV
provoked great theoretical interest because the comparatively large cross
section revealed the influence of unexpected short-range physics \cite{Riska}.
 
The total and differential cross sections for charged pion production in the
allied $pp\to pn\pi^+$ reaction have also been measured very close to
threshold \cite{Daehnick1,Daehnick2} and these results have recently been
complemented by the determination of the proton analyzing power
\cite{Flammang}.  At the same value of $Q$, charged pion production is
typically a factor of five larger than neutral and the interesting
practical question to ask is to what extent the extra information will
constrain theory further.

The general features of the first results on the $pp\to pn\pi^+$ total
cross section near threshold \cite{Daehnick1} can in fact be understood within
a nucleon-nucleon final state interaction model using measurements of the
$pp\to d\pi^+$ and $pp\to pp\pi^0$ reactions as input \cite{FW5}. This approach
uses an approximation for the $np$ spin-triplet scattering wave function in
terms of the deuteron bound-state wave function which does not depend upon the
details of the $np$ potential \cite{FW4,FW8}.  This result, which is exact when
the $np$ scattering energy is extrapolated to $-\epsilon_d$,  where
$\epsilon_d$ is the deuteron binding energy, is generally a very good
approximation for small energies and distances \cite{Paris}.  Taking the
pion production operator to be of short range, the formalism allows one to
estimate the cross section for the $pp\to pn\pi^+$ reaction where the final
$np$ pair are in the spin-triplet $S$-wave. The much smaller contribution from
spin-singlet final states may be deduced directly from the $pp\to pp\pi^0$ data
using isospin invariance.

The agreement with the measured total cross sections is generally very
encouraging except for underestimating the point nearest threshold \cite{FW5},
which is the hardest to measure experimentally. In the definitive data set of
Ref.\cite{Daehnick2}, the value at this point is somewhat reduced and a
more careful evaluation of Coulomb effects given in this present work 
increases the near-threshold prediction. These two changes reduce 
the discrepancy significantly.

Away from threshold our final state interaction approach is capable of
describing the angular distribution of the TRIUMF $pp\to pn\pi^+$ data at 400
and 450~MeV \cite{Falk} provided the excitation energy in the $np$ final state
is below about 10~MeV, where $S$-waves can be expected to dominate \cite{FW6}.
Differential cross sections in the pion c.m.~angle are now also available near
threshold \cite{Daehnick2}, and we here extend our formalism to describe such
data. A major uncertainty in implementing this program is the restricted
differential cross section information for the $pp\to pp\pi^0$ reaction
corresponding to the total cross sections reported in Ref.\cite{Meyer}.
Assuming this to be isotropic, we predict the $pp\to pn\pi^+$ angular
distribution to be a little steeper than that observed experimentally
\cite{Daehnick2}, though it should be noted that there are also 
uncertainties in the $pp\to d\pi^+$ input \cite{Arndt,pppid1,pppid2}. 

The proton analyzing power in low energy $\vec{p}p\to d\pi^+$ is largest at
$90^0$ and the energy dependence of this quantity for the three-body 
$pp\to pn\pi^+$ final state has just been reported by an IUCF group
\cite{Flammang}. The extension of our FSI method to treat analyzing powers is
straightforward, but its application is hindered by the dearth of $A_y$ data 
in the $\vec{p}p\to pp\pi^0$ spin-singlet final state. Assuming this to 
be small, the broad features of the $90^0$ energy dependence are reproduced.

Calculations within our final state interaction model are not definitive in
that they do not use all of the energy dependence of the $S$-state
neutron-proton scattering wave functions. This means however that they
do not require detailed considerations of the range dependence of the 
pion production operator. Higher partial waves in the neutron-proton 
final state are simply ignored and this will become a bigger defect with 
increasing energy. Nevertheless we claim that any microscopic calculation 
which reproduces the measured $pp\to d\pi^+$ and $pp\to pp\pi^0$ 
observables should, near threshold, give results qualitatively similar
to the ones presented here. The tolerable agreement with the $pp\to pn\pi^+$
data therefore suggests that the extraction of quantitatively new information 
on the basic pion production reaction mechanism near threshold requires 
accurate measurements combined with much more refined theoretical calculations.

In Sec.\ 2 we discuss the relation of the $NN$ scattering and bound state wave
functions in the spin-triplet $S$-state and illustrate it with the results
obtained from the Paris potential \cite{Paris}. The energy dependence of the
spin-singlet wave functions is investigated with and without Coulomb effects
and it is shown that the major difference between the two can be accounted for
by merely shifting the position of the $^1$S$_0$ pole on the second sheet.

Though at low energies it is reasonable to consider only $S$-wave 
nucleon-nucleon systems, pionic $p$-waves enter quickly because of the
smallness of the $s$-wave terms. In Sec.\ 3 we develop the formalism to
include the effects of pionic $s$- and $p$-waves, together with the 
approximate wave 
functions, in the three-body phase space to estimate the measured 
$pp\to pn\pi^+$ observables. Coulomb corrections are included in the 
$s$-wave in the same approximate manner as that used to treat the 
low energy $pp\to d\pi^+$ reaction.

As previously stressed, the agreement with the experimental data presented in
Sec.\ 4 is generally good, though not perfect. However the differences are well
within the uncertainties of the model and the input and output experiments.
Conclusions and suggestions for further work are to be found in Sec.\ 5.
\protect\section{Scattering and Bound-State Wave Functions}

An $S$-wave bound state reduced wave function behaves asymptotically as
\begin{equation}
\label{2_1}
u_{\alpha}(r) \sim e^{-\alpha r}
\end{equation}
at large radii $r$ and is normalised such that
\begin{equation}
\label{2_2}
\int_{0}^{\infty}dr\, [u_{\alpha}(r)]^2 =1  \: . 
\end{equation}

The analogous scattering wave function with {\it real} boundary conditions is 
normalised by its asymptotic behaviour
\begin{equation}
\label{2_3}
v(k,r)\sim\frac{1}{k}\sin (kr+\delta(k)) \: ,
\end{equation}
where $\delta(k)$ is the phase shift at wave number $k$. 

Despite these very different normalisations, we have shown that for finite
range potentials the two types of wave functions are related quantitatively by
the theorem \cite{FW8}
\begin{equation}
\label{2_4}
\lim_{k\to i\alpha}\left\{\sqrt{2\alpha(\alpha^2+k^2)}\,v(k,r)\right\}= 
-u_{\alpha}(r) \:.              
\end{equation}

This theorem is valid at the bound state pole, independent of the details of
the potential, but it is also a robust extrapolation
in the sense that, in the case of a weakly bound state, at short distances $r$
\begin{equation}
\label{2_5}
v(k,r) \approx -\frac{1}{\sqrt{2\alpha(\alpha^2+k^2)}}\,u_{\alpha}(r) \:, 
\end{equation}
provided that $k$ is small on the scale of the potential range and strength.
Of course the approximation breaks down at large distances because the
bound and scattering wave functions must be orthogonal when integrated
over all space.

In the case of an almost bound virtual state, where the pole is on the second 
sheet, there is no bound state wave function to fix the radial dependence. 
Nevertheless the same techniques show that the energy dependence of the 
scattering wave function is given by a similar factor to that of 
Eq.\ (\ref{2_5}) \cite{FW5,FW8}: 
\begin{equation}
\label{2_6}
v(k,r) \approx -\frac{1}{\sqrt{\alpha^2+k^2}}\,C(r) \:.
\end{equation}

These approximations can be tested explicitly in the cases of the Yamaguchi,
square well, Bargmann or Hulth\'{e}n potentials, which can be resolved exactly.
However several complications arise when looking at more realistic potentials
which describe the nucleon-nucleon system. The extrapolation theorem of
Eq.\ (\ref{2_5}) has only been proved for single channel scattering in a local
potential \cite{FW8}, whereas in the spin-triplet case there is coupling 
between the $S$ and $D$-waves. 

In Ref.\cite{FW5} we showed the variation of the deuteron and scattering 
wave functions with $r$ and, despite the neglect of coupled channel effects,
Eq.\ (\ref{2_6}) seems to be broadly valid to within 5\% for $r\leq 1.7$~fm 
for neutron-proton c.m.~energies $E_{np}\leq 10$~MeV. To see this more
quantitatively, we plot in fig.~1 values of the spin-triplet function
\begin{equation}
\label{2_7}
Z_t=-\sqrt{2\alpha_t(k^2+\alpha_t^{\,2})}\:v(k,r)/r\:,
\end{equation}
with $\alpha_t=0.232$~fm$^{-1}$,
in terms of $k^2$ \cite{Paris}. The radius is fixed at $r=1.05$~fm, which is
close to the peak in the Paris wave function.

The smooth fit
\begin{equation}
\label{2_8}
Z_t(k^2) = \sqrt{2\alpha_t}\,(0.627+0.122k^2-0.057k^4)
\end{equation}
lies 0.5\% below the value of the deuteron wave function, but that is close to
the limit in the precision of the extrapolation. This deviation could of course
be due to effects of the $D$-state, though the Paris potential also includes 
some velocity dependence which may lie outside the domain of validity of the
theorem. While almost all the energy dependence is given by the  square-root
factor in Eq.\ (\ref{2_6}),  $Z_t(k^2)$ is a steadily increasing function and
this is typical for potentials with just one lightly bound state. 
Nevertheless, even at $E_{np}=15$~MeV it lies less than 8\% above the deuteron
point.  It should be noted that the slope of this function depends  upon the
value of $r$ and that at $r\approx 1.7$~fm the function is almost flat.

Though there are no channel-coupling problems in the spin-singlet case, 
the approximation given by Eq.\ (\ref{2_6}) must break down at very small $k^2$
in the proton-proton case due to the essential singularity at $k=0$ caused by
the Coulomb repulsion. This is not a problem provided $E_{NN}>1$~MeV, as can
be seen from the values of the singlet function 
\begin{equation}
\label{2_9}
Z_s=-\sqrt{k^2+\alpha_s^2}\:v(k,r)/r
\end{equation}
which are also shown in fig.~1.

The results may be parametrised by
\begin{eqnarray}
\nonumber
Z_s=0.646+0.343k^2-0.353k^4+0.157k^6\,,\ \ 
&& \text{with } \alpha_s=-0.100\,\mbox{\rm fm}^{-1}\ \ 
\text{(without Coulomb),}\\
\label{2_10}
Z_s=0.698+0.131k^2-0.037k^4\,,\ \ \
&&\text{with } \alpha_s=-0.053\,\mbox{\rm fm}^{-1}\ \ \text{(with Coulomb).}
\end{eqnarray}
\protect\section{Amplitudes and Observables}

Many of the elements of the FSI formalism are given in Ref.\cite{FW5}, but for
clarity some are repeated here. We use a normalisation such that the
two-body spin-averaged amplitude and c.m.~differential cross section are 
related by
\begin{equation}
\label{3_1}
\frac{d\sigma}{d\Omega}(pp\to d\pi^+)=
 \frac{p_{\pi}}{64\pi^2(m_d+m_{\pi})^2p_p}
\times |M(pp\to d\pi^+)|^2\:,
\end{equation}
where $m_p$, $m_{n}$, $m_{d}$, and $m_{\pi}$ are the masses of the proton,
neutron, deuteron, and pion, and $p_{\pi}$ and $p_p$ are the momenta of the 
final pion and the initial proton respectively.

The corresponding three-body cross section is given by
\begin{eqnarray}
\label{3_2}
\frac{d^2\sigma}{d\Omega\,dk}(pp\to pn\pi^+)&=& 
\frac{p_{\pi}\,k^{2}}{32\pi^{3}(m_p+m_n+m_{\pi})^2 p_p} 
\times |M(pp\to (pn)_{k}\pi^+)|^2\:,
\end{eqnarray}
where the matrix element squared is assumed to be averaged over the angles of
the $np$ relative momentum $\vec{k}$. In practice we shall only estimate the
contributions from $S$-wave $np$ pairs, where the results are in any case
isotropic in $\vec{k}$.

Reduced mass factors for the three-body reaction are defined by
\begin{equation}
\label{3_4}
\frac{1}{\mu} = \frac{1}{m_{\pi}} + \frac{1}{m_p+m_n}\:;\ \ \ \ \ \ \ 
\frac{1}{m_{pn}}= \frac{1}{m_p} + \frac{1}{m_n}\:,
\end{equation}
whereas for bound state production the $m_p+m_n$ combination in 
Eq.\ (\ref{3_4}) is replaced by the deuteron mass.

The momenta in the exit channel are linked to the value of the excitation
energy $Q$ through
\begin{equation}
\label{3_5}
Q=\frac{p_{\pi}^{\,2}}{2\mu}+\frac{k^2}{2m_{pn}}\:\cdot
\end{equation}
The variable $\eta$, conventionally  used to describe low energy pion 
production, is the maximum pion momentum in pion mass units and is 
given by
\begin{equation}
\label{3_6}
\eta \equiv \frac{p_{\pi}^{max}}{m_{\pi}} = \frac{\sqrt{2\mu Q}}{m_{\pi}}
\:\cdot
\end{equation}

The cross section for pion production summed over all $pn$ excitation energies
in the three-body channel is
\begin{eqnarray}
\label{3_7}
\frac{d\sigma}{d\Omega}(pp\to &&pn\pi^+)= 
\frac{1}{32\pi^{3}(m_p+m_n+m_{\pi})^2 p_p}\:
\int_{0}^{\sqrt{2m_{pn}Q}}
|M(pp\to (pn)_{k}\pi^+)|^2\, p_{\pi}\, k^{2}\,dk\:.
\end{eqnarray}

Since the pion production operator is expected to be dominated by short-range
physics, the relation of Eq.\ (\ref{2_4}) between the bound and scattering wave
functions suggests a similar relation for the spin-triplet matrix elements,
viz
\begin{eqnarray}
\label{3_8}
|M(pp\to (pn)_t\pi^+)|^2 \approx \frac{|M(pp\to d\pi^+)|^2}
{2\alpha_t(k^2+\alpha_t^{\,2})}\approx
\frac{32\pi^2(m_d+m_{\pi})^2p_p}{p_{\pi}\alpha_t(k^2+\alpha_t^{\,2})}\:
\frac{d\sigma}{d\Omega}(pp\to d\pi^+)\:.
\end{eqnarray}
The prediction for the spin-triplet part of the unpolarised cross section then
follows immediately through the use of Eq.\ (\ref{3_7}). 

The formalism for the
proton analyzing power $A_y$ in $pp\to pn\pi^+$ can be developed identically,
the only changes being $A_y$ factors on the right hand side of Eq.\ (\ref{3_8})
and the left hand side of Eq.\ (\ref{3_7}).

It is of great practical interest that near threshold the integration in
Eq.\ (\ref{3_7}) may be performed analytically. Thus for a $pp\to d\pi^+$ 
observable which varies as
\begin{equation}
{\cal O}_{d}^{(n)}(\eta) \propto \eta^n\:,
\end{equation}
the corresponding observable for the $pp\to (pn)_t\pi^+$ reaction is given
by
\begin{equation}
\label{3_9}
{\cal O}_{pn}^{(n)}(\eta) = P^{(n)}(\eta\zeta) \times 
{\cal O}_{d}^{(n)}(\eta)\:,
\end{equation}
where
\begin{equation}
\label{3_10}
\zeta= \frac{m_{\pi}}{\sqrt{2\mu\epsilon}}\:\cdot
\end{equation}

\newpage
A description of the low energy $pp\to pn\pi^+$ observables only requires these
functions for $n=1,\, 2,\, 3$, and explicitly in these cases

\begin{mathletters}
\label{3_11}
\begin{eqnarray}
P^{(1)}(x)&=&\frac{x^3}{4(1+\sqrt{1+x^2})^{2}}\:,\\[1ex]
P^{(2)}(x)&=&\frac{1}{\pi}\left\{\frac{2}{3}x + \frac{1}{x} 
-\left(1+\frac{1}{x^2}\right)\arctan(x)\right\}\:,\\[1ex]
P^{(3)}(x)&=& \frac{x^3}{8(1+\sqrt{1+x^2})^{2}}
\left\{1+\frac{x^2}{2(1+\sqrt{1+x^2})^{2}}\,
\right\}\,\cdot
\end{eqnarray}
\end{mathletters}

It is customary in the analysis of  $pp\to d\pi^+$ data to assume that, due to
the Coulomb repulsion in the final state,  the cross section very near
threshold is suppressed by a Gamow factor of
\begin{equation}
\label{3_12}
F_C(\eta) 
=\frac{2q/\eta}{\exp(2q/\eta)-1}
\approx \frac{1}{1+q/\eta}
\:,
\end{equation}
where
\begin{equation}
\label{3_13}
q=\frac{\pi\mu\alpha}{m_{\pi}}\:\cdot
\end{equation}

This approach, which considers all the charge in the deuteron to be
concentrated at the center of the nucleus, is sufficient to explain the major
differences between the $np\to d\pi^0$ \cite{Hutcheon} and $pp\to d\pi^+$
\cite{Arndt} total cross sections, and the simplification made in
Eq.\ (\ref{3_12}) is indistinguishable from the full Gamow factor under the
conditions of such data.

An analogous suppression will be suffered by the $\pi^+$ in the $pp\to pn\pi^+$
case, but this is only likely to be of practical importance for $s$-wave pions
because the  centrifugal barrier keeps the $p$-waves small in the
region of low
$\eta$.  Taking the charge, as in the deuteron case, to be situated at the
center of mass of the proton-neutron  system, the integral for the 
Coulomb-modified $P^{(1)}(x)$ of Eq.\ (\ref{3_11}) can still be performed in 
closed form, giving

\newpage
\[P_{C}^{(1)}(x,c)=
\frac{x^3}{\pi(1+x^2(1-c^2))}\,\left\{\mbox{${\textstyle \frac{1}{2}}$}
 c^3\sqrt{1-c^2}\,\ell n\left(\frac{1-\sqrt{1-c^2}}{1+\sqrt{1-c^2}}\right)
\right.\]
\[+\frac{\pi}{4x^4}(1+x^2(1-c^2))(2+x^2(1+2c^2))
-\frac{\pi}{2x^4}\,(1+x^2)^{3/2}\]
\begin{equation}
\label{3_14}
\left.-\frac{c}{x^2}(1+x^2(1-c^2))+
\frac{c}{x^3}\,(1+x^2)\arctan(x)\right\}\:,
\end{equation}
where $c=q/\eta$.

In $pp \to pp\pi^0$ the $s$-wave pion production is likely to be dominant to 
much higher values of $\eta$ than for charged pion production induced by the 
suppression of the $\Delta$ contributions through the selection rules
for the final $S$-wave $pp$ states. 
Though there is no bound state production to set the normalisation, 
the energy dependence of the total cross section should therefore be of
the form $\eta P^{(1)}(x)$. In the absence of Coulomb effects, the cross
sections for $\pi^0$ and $\pi^+$ production are related by
\begin{equation}
\label{3_15}
\sigma(pp\to \{pn\}_{I=1}\pi^+) = \sigma(pp\to pp\pi^0)\:.
\end{equation}

Coulomb corrections must however be applied before this is used to
estimate the spin-singlet contribution to $pp\to pn\pi^+$. On the
right hand side the singlet pole in the wave function is shifted
as in Eq.\ (\ref{2_10}), whereas on the left the Coulomb modified 
$P_{C}^{(1)}(x,c)$ is used instead of $P^{(1)}(x)$. 

\protect\section{Comparison with Experiment} 

In the low energy region it is a good approximation to take the $pp\to d\pi^+$
differential cross section and analyzing power to be of the form
\begin{mathletters}
\label{4_1}
\begin{eqnarray}
\frac{d\sigma}{d\Omega} &=& A + B\cos^2\theta\:,\\
A_y\,\frac{d\sigma}{d\Omega} &=& C\sin\theta\:,
\end{eqnarray}
\end{mathletters}
where $\theta$ is the c.m.~angle of the pion,
and these dependences will propagate through to the spin-triplet part of the 
$pp\to pn\pi^+$ reaction.

The SP96 phase shift solution from the Virginia database \cite{Arndt} yields 
the following Coulomb-corrected predictions for observables. At $\theta=0^0$, 
\begin{equation}
\label{4_2}
\frac{d\sigma}{d\Omega^*} =(13.6\,\eta + 164.0\,\eta^3)\ \mu\text{b/sr}\:,
\end{equation} 
whereas at $\theta=90^0$ 
\begin{mathletters}
\label{4_34}
\begin{eqnarray}
\label{4_3} 
\frac{d\sigma}{d\Omega^*}
&=&(13.6\,\eta + 16.2\,\eta^3)\ \mu\text{b/sr}\:,\\[1ex]
\label{4_4}
A_y\,\frac{d\sigma}{d\Omega^*} &=&-18.1\,\eta^2 \ \mu\text{b/sr}\:, 
\end{eqnarray}
\end{mathletters}
though it should be noted that the $s$-wave strength reported here is 7\% less
than some direct measurements \cite{pppid1,Hutcheon}.

Spin-singlet final states may be estimated using
\begin{equation}
\label{4_5}
\sigma(pp\to pp\pi^0)\approx 4\pi D\,\eta\,P^{(1)}(\zeta'\eta)
\end{equation}
where, as discussed in Sec.\ 2, after including the $pp$ Coulomb repulsion, 
$\zeta' =12.6$.
The best fit to the total cross section data of Ref.\cite{Meyer} is achieved 
with $D\approx 0.86\ \mu$b, and the resulting
curve shown in fig.~2 is very similar in 
shape to microscopic calculations which include the $pp$ final state
interaction correctly \cite{Hor}.

Though the authors of Ref.\cite{Meyer} did not report values for the
$pp\to pp\pi^0$ differential cross sections, their data do suggest
some influence from higher partial waves at 325~MeV. 
There is however evidence that at 320~MeV ($\eta=0.46$) the cross 
section is fairly isotropic and that the proton analyzing power is
consistent with zero \cite{Stan}. 
Introducing Coulomb corrections in both the final $pp\pi^0$ and
$pp\pi^+$ systems, these data then predict that
\begin{equation}
\label{4_6}
\frac{d\sigma}{d\Omega^*}(pp\to \{pn\}_s\pi^+)
\approx D'\,\eta\,P_{C}^{(1)}(\zeta''\eta,c)\:,
\end{equation}
where $D'\approx 0.42\ \mu$b/sr and $\zeta''=25.4$. This is to be taken in
conjunction 
with the spin-triplet predictions for the differential cross section 
\begin{eqnarray}
\label{4_7}
\frac{d\sigma}{d\Omega^*}(pp\to \{pn\}_t\pi^+)
=13.6\,\eta\,P^{(1)}_{C}(\zeta''\eta,c)
+ (16+ 148\cos^2\theta)\,\eta^3\,P^{(3)}(\zeta''\eta)\ \mu\text{b/sr}\:,
\end{eqnarray} 
and analyzing power
\begin{equation}
\label{4_8}
A_y\,\frac{d\sigma}{d\Omega^*}(pp\to \{pn\}_t\pi^+)
 =-18\,\eta^2\sin\theta\,P^{(2)}(\zeta''\eta)\:  \mu\text{b/sr}\:,
\end{equation} 
where pion Coulomb corrections have only been made in the $s$-wave.

The momentum dependence of the $pp\to pn\pi^+$ total cross section is shown in
fig.~2 both with and without pionic Coulomb corrections. There is in fact very
little difference between either using the $P_{C}^{(1)}(x,c)$ of 
Eq.\ (\ref{3_14}) or
multiplying the uncorrected $P^{(1)}(x)$ of Eq.\ (\ref{3_11}) by the 
Gamow factor of Eq.\ (\ref{3_12}). The full curve lies somewhat below 
the IUCF measurements \cite{Daehnick2} but, apart from the lowest point, 
the deviations are of the same order as the Coulomb differences. It should 
be noted that there are typically 10\% overall normalisation uncertainties 
in both the input and output data, as well as uncertainties in determining 
the value of $\eta$ close to threshold \cite{Daehnick2}. The total
cross sections from the extended data set of Ref.~\cite{Flammang} are
consistent with the earlier measurements, though they are subject to
very similar systematic uncertainties, 

Due to the differences in the pole positions, the ratio of the
$s$-wave production of pions with spin-triplet and spin-singlet final 
states is estimated to be about $1.7$ at low $\eta$ but $7.8$ at high $\eta$. 

The angular dependence of the predicted $pp\to pn\pi^+$ differential cross 
section at $\eta=0.21$ and $\eta=0.42$ is compared in fig.~3 with the IUCF data
\cite{Daehnick2}. The theory is about 20\% too low in normalization
and the slopes, evaluated
assuming the singlet cross section to be isotropic, a little too
strong. It is clear from the larger data set of Ref.~\cite{Flammang}
that the binning effect due to the experimental averaging over angular 
domains is only of marginal importance.

In Fig.~4 is shown the variation of $A_y$ at $90^0$ as a function of $\eta$ for
$pp\to d\pi^+$ and $pp\to pn\pi^+$, assuming the singlet cross section to be
isotropic with zero analyzing power. The curve lies a little below the
experimental points of Ref.\cite{Flammang}, where the only major systematic
uncertainty is the acceptance in polar and azimuthal angles.

\protect\section{Conclusions}

We have shown within the framework of a simple final state interaction theory
that most of the low energy data on the total and differential cross sections
and proton analyzing power for the $pp\to pn\pi^+$ reaction can be understood
semi-quantitatively in terms of equivalent information deduced from $pp\to
pp\pi^0$ and $pp\to d\pi^+$ measurements. We would argue that any more
microscopic approach with finite range pion-production operators, which fit 
the latter data, would give $pp\to pn\pi^+$ predictions broadly similar to the
ones presented here, though with some deviations at higher values of
$\eta$ due to the neglect of final $np$ $P$-wave contributions. 
Such contributions might be responsible for some of the discrepancies
observed in the angular distributions and analyzing power.
To take full advantage of the accurate IUCF data
would require a sophisticated microscopic model to be accurate to the
10\% level. It could then be helpful if such a model were
consistently applied in the evaluation of the experimental acceptances
rather than relying on the Watson FSI approach \cite{Daehnick2}.

Qualitatively new information, against which to test our approach,  
would be provided by experiments
which introduce interferences between triplet and singlet final $np$ 
states, as for example in the $(\vec{p},\vec{p}\,')$
spin-transfer. These are however much harder to perform in practice.

The deviations between our predictions and the IUCF data
\cite{Daehnick1,Daehnick2,Flammang} may be partially experimental in origin.  It
should be noted that there are uncertainties of the order of 10\% in some of the
$pp\to d\pi^+$ observables used as input, and this is worse for the differential
quantities in $pp\to pp\pi^0$. However  it is also apparent from fig.~1 that our
formalism  will tend to  underestimate slightly the  triplet contribution to the
$pp\to pn\pi^+$ cross section because $Z_t$ is an increasing function of  $k^2$.
The rate  of rise  depends  upon the   nucleon-nucleon  separation  and so it is
impossible  to  quantify the  effect  without  having a  detailed  model for the
pion-production    operator.   This  is   therefore left  for  more  microscopic
formulations.

\acknowledgements
We are very grateful to W.W.~Daehnick for supplying us with the analyzing 
power data reported in Ref.\cite{Flammang} prior to publication, and for 
explaining carefully the significant points of this and the unpolarised 
cross sections. We should also like to thank R.W.~Flammang for the
detailed information contained in his thesis. Useful correspondence 
with H.O.~Meyer and W.W.~Jacobs is also acknowledged. This work has
been made possible by the continued financial support of the Swedish Royal
Academy and one of the authors (C.W.) would like to thank them and the The
Svedberg Laboratory for their generous hospitality.

\baselineskip 3ex
\newpage
\input epsf
\begin{figure}[t]
\begin{center}
\mbox{\epsfxsize=5in \epsfbox{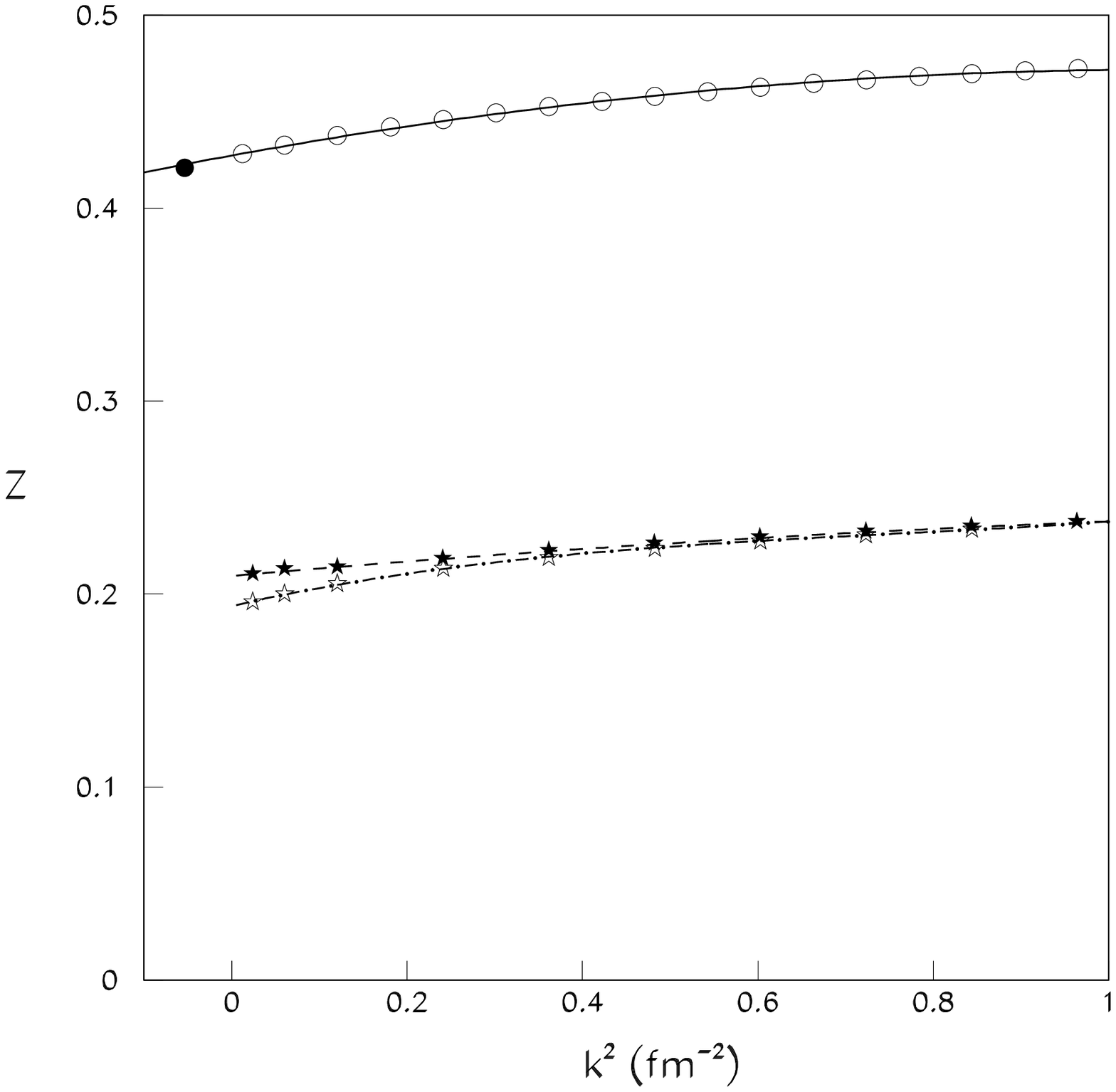}}
\end{center}
\end{figure}
\noindent Figure 1:
The NN triplet scattering function $Z_t$ defined by
Eq.\ (\ref{2_7}) evaluated at $r=1.05$~fm for discrete values of 
$k^2$ using the Paris potential (open circles). The smooth extrapolation 
of Eq.\ (\ref{2_8}) lies slightly above the
deuteron point (closed circle). The points for the corresponding singlet 
functions $Z_s$, defined by Eq.\ (\ref{2_9}), are shown without Coulomb force 
(open star) and with (closed star) and interpolated using Eq.\ (\ref{2_10}). 
The principal effect of the Coulomb force is to change the position of the
antibound state pole at $k^2=-\alpha_s^{\,2}$. Residual effects due to the
Coulomb repulsion may be seen below $E_{pp}=1$~MeV.

\newpage
\input epsf
\begin{figure}[t]
\begin{center}
\mbox{\epsfxsize=5in \epsfbox{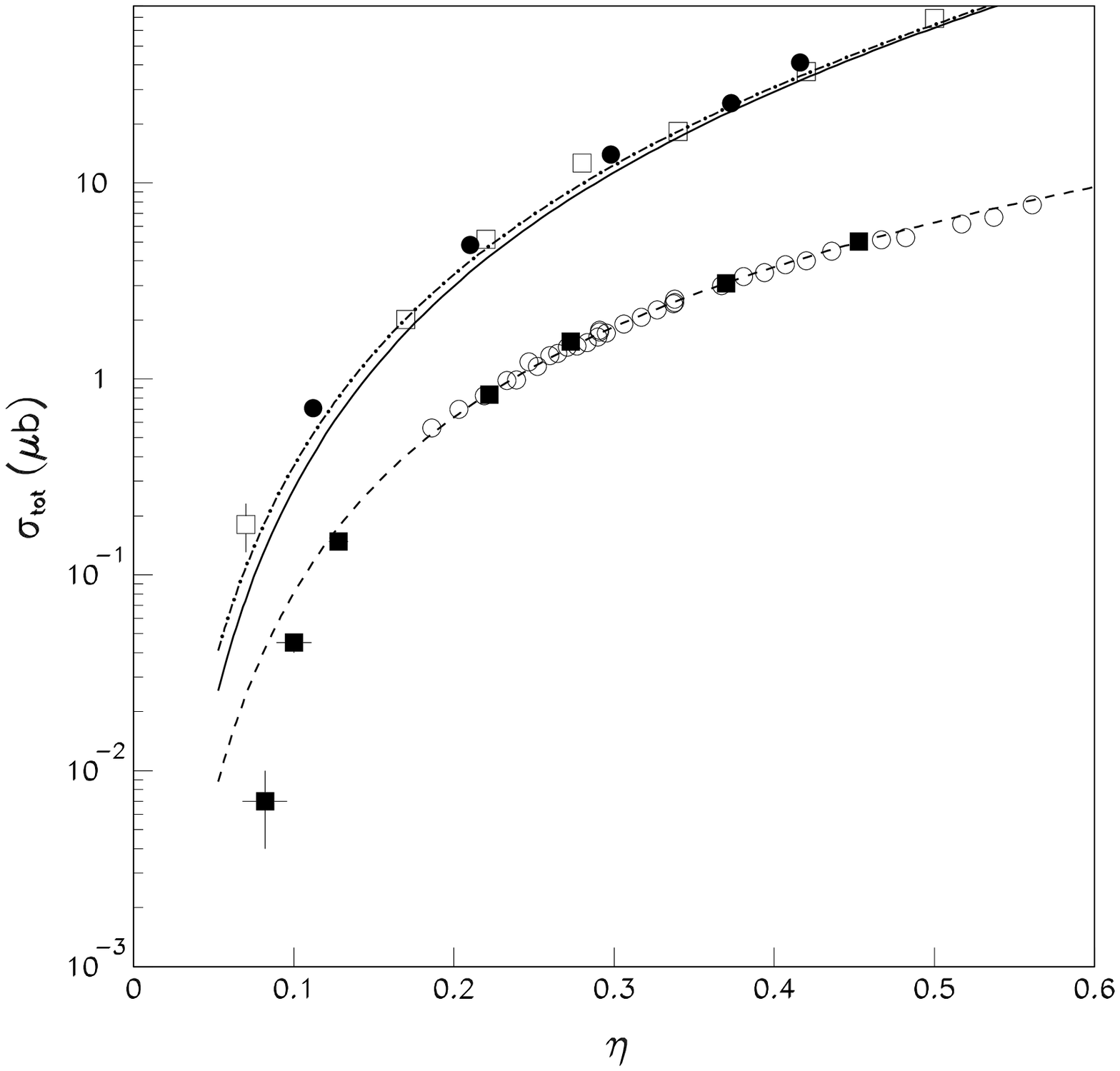}}
\end{center}
\end{figure}

\noindent
Figure 2:
The experimental $pp\to pp\pi^0$ total cross sections of 
Ref.\protect\cite{Meyer} (open circles) and Ref.\protect\cite{Bondar} 
(squares) are to be compared to the broken curve, which represents the
fit on the basis of Eq.\ (\protect\ref{4_5}). This curve is very 
similar in shape to the microscopic calculations of 
Refs.\protect\cite{Hor}. The $pp\to pn\pi^+$ data of 
Ref.\protect\cite{Daehnick2}, which are very similar to the extended 
data set of Ref.~\protect\cite{Flammang}, are subject to a 13\% 
overall uncertainty. The solid and chain curves show the corresponding
predictions with and without the pionic Coulomb corrections. The
Coulomb corrections given by Eq.\ (3.14) differ little in practice from 
those obtained by merely multiplying by the Gamow factor of Eq.\ (3.12).

\newpage
\input epsf
\begin{figure}[t]
\begin{center}
\mbox{\epsfxsize=5in \epsfbox{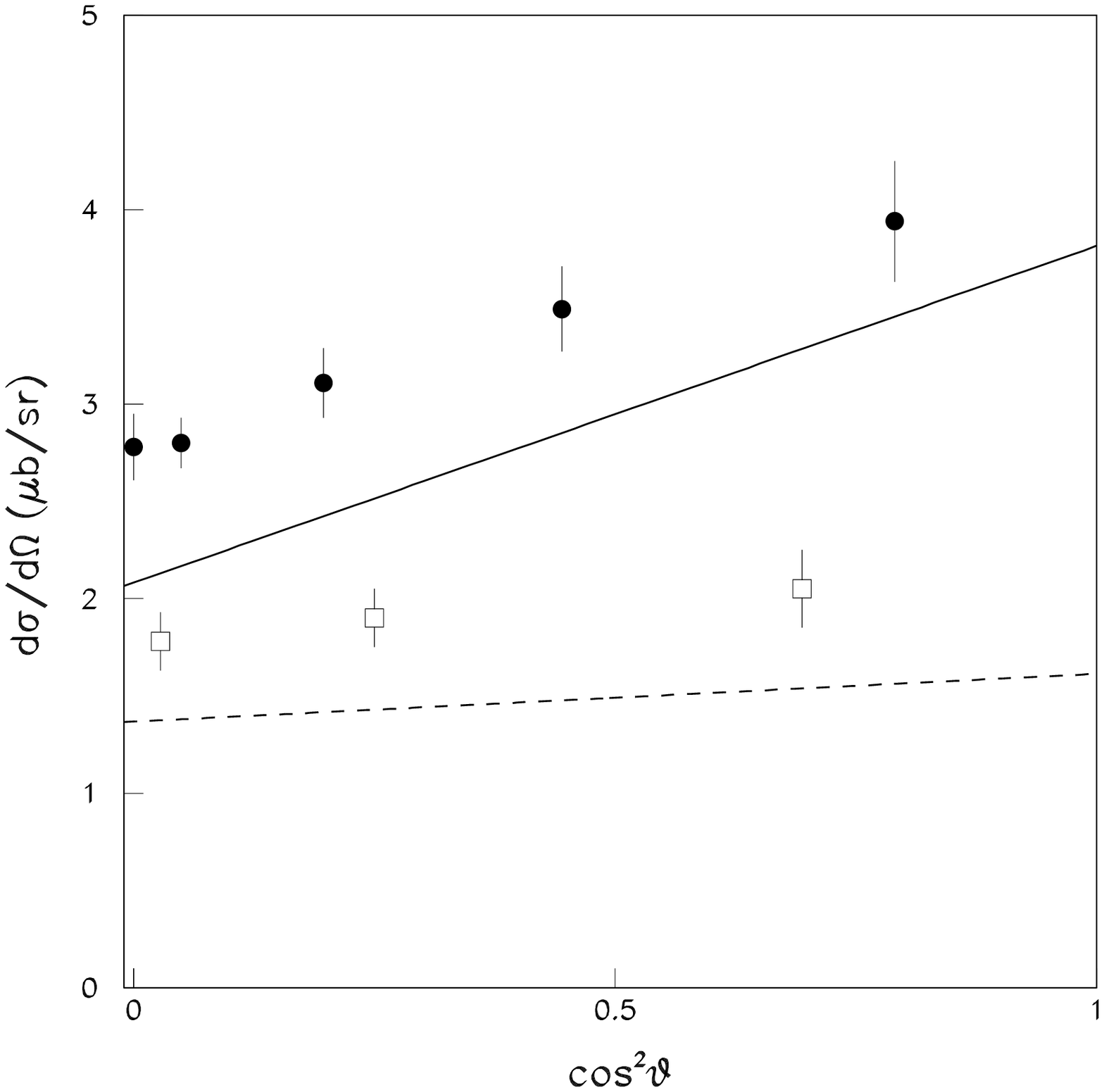}}
\end{center}
\end{figure}

\noindent
Figure 3:
Differential cross sections for $pp\to pn\pi^+$ at $\eta=0.21$ (squares) and 
$\eta=0.42$ (circles) taken from Ref.\protect\cite{Daehnick2} and 
compared with the
predictions of the model assuming the singlet cross section to be isotropic.
The data and predictions at the lower energy are both multiplied by a factor of
five to present them on a similar scale.

\newpage
\input epsf
\begin{figure}[t]
\begin{center}
\mbox{\epsfxsize=5in \epsfbox{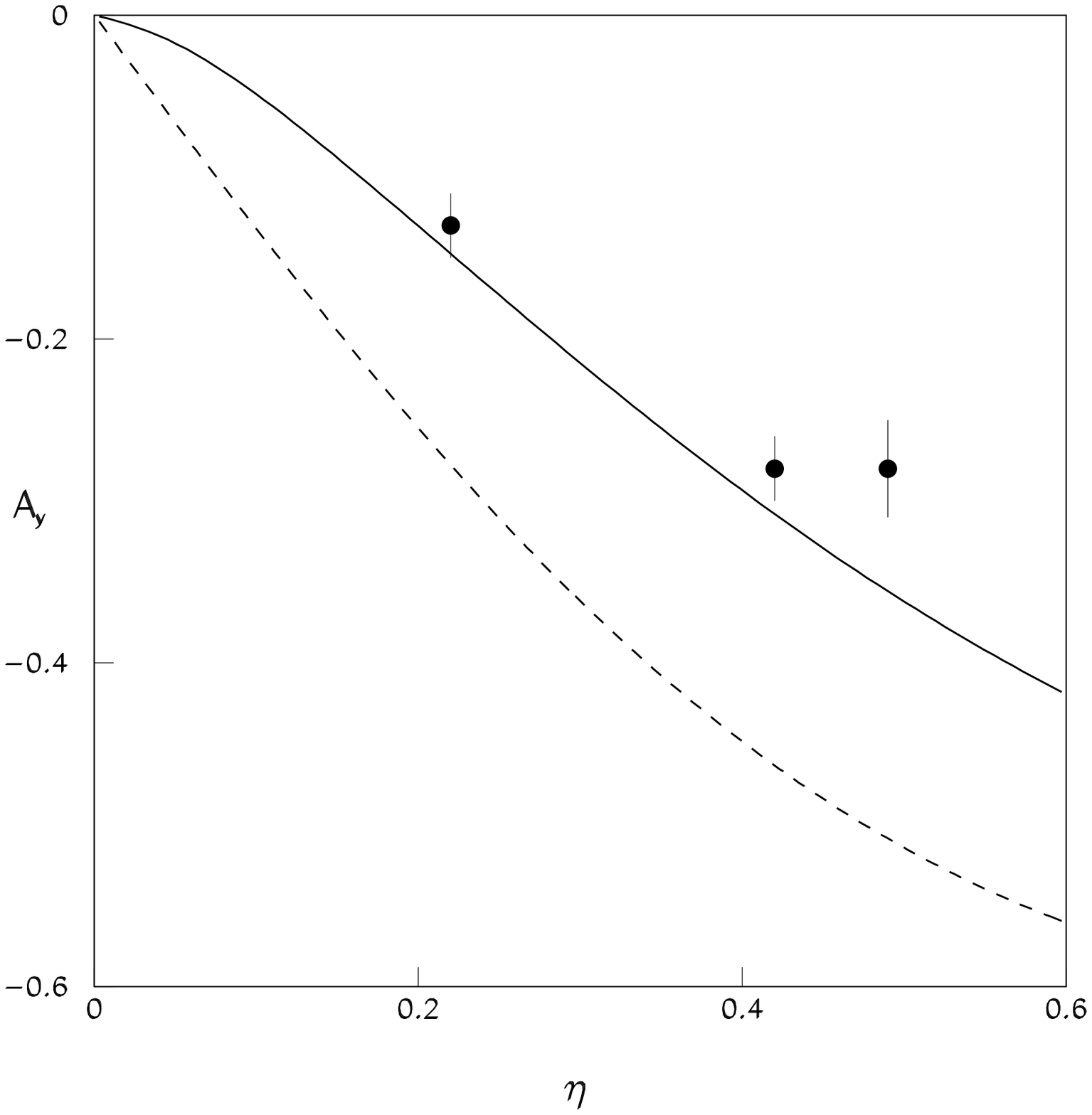}}
\end{center}
\end{figure}

\noindent
Figure 4:
Variation of the proton analyzing power in $pp\to pn\pi^+$ at $\theta=90^0$ as 
a function of $\eta$, the experimental points being taken from
Ref.\protect\cite{Flammang}. The broken curve is the prediction 
of the phase shift solution SP96 of Ref.\protect\cite{Arndt} 
for $pp\to d\pi^+$, whereas the solid curve
follows from Eq.\ (\ref{4_8}) assuming the $pp\to pp\pi^0$ cross section to be
isotropic with negligible analyzing power.
\end{document}